\documentstyle[11pt,psfig,wrapft]{article}
\begin{document}

\begin{titlepage}
\begin{flushright}
CERN-TH/2001-070\\
HUPD-0102
\end{flushright}

\begin{center}
\vspace*{1.6cm}

{\LARGE\bf Large-$N_c$ meson theory}
\vspace{1.2cm}

Shin-Ichiro~Kuroki
\footnote{E-mail address: {\tt kuroki@theo.phys.sci.hiroshima-u.ac.jp}},
Keiichi~Morikawa
\footnote{E-mail address: {\tt morikawa@sxsys.hiroshima-u.ac.jp}} and\ \   
Takuya~Morozumi
\footnote{On leave of absence from Hiroshima University. E-mail address: {\tt takuya.morozumi@cern.ch}}
\vspace{8mm}

$^1$ {\it Graduate School of Science, Hiroshima University, 1-3-1 Kagamiyama, Higashi-Hiroshima, Japan}\\ \vspace{3mm}
$^2$ {\it  Research Center for Nanodevices and Systems, Hiroshima University, 1-4-2~ Kagamiyama, Higashi-Hiroshima, Japan}\\ \vspace{3mm}
$^3$ {\it  CERN, Geneva, Switzerland}
\vspace{1.8cm}

\begin{abstract}
We derive an effective Lagrangian for meson fields.
This is done in the light-cone gauge for two-dimensional large-$N_c$ QCD
by using the bilocal auxiliary field method.
The auxiliary fields are bilocal on light-cone space and their Fourier
transformation determines the parton momentum distribution.
As the first test of our method, the 't~Hooft equation is derived
from the effective Lagrangian.
\end{abstract}

\end{center}
\end{titlepage}

\newpage
\section{Introduction}

 QCD successfully describes the dynamics of quarks and gluons in the
 perturbative regime. However the dynamics of mesons and baryons are
 not yet understood from the first principle. Although we believe that
 the meson is the bound state of quarks and the numerical
 simulation based on lattice QCD is successful, it still is a difficult
 task to clarify this connection analytically. Large-$N_c$ QCD has
 played an important role on our understanding of
 QCD [1--3]. Especially in two
 dimensions, that is, in the 't Hooft model, a bound-state equation
 is obtained. This is the well-known 't Hooft equation~\cite{thooft,coleman},
 and it is derived from a ladder Bethe--Salpeter equation. 
 This equation can be solved by several methods, e.g., the discretized
 light-cone quantization~\cite{brodsky-1,brodsky-2}, the variational
 method~\cite{harada}, and the Multhopp technique~\cite{grins}. 
 These calculations give us various information about the meson wave
 function and the bound state mass. Our problem is how we can
 understand this equation in the Lagrangian level and, more, how we can
 deduce the Lagrangian corresponding to this equation from QCD.

 In this paper, we derive an effective Lagrangian of large-$N_c$
 QCD$_2$. The derivation of the Lagrangian is based on the bilocal
 auxiliary field method [8--10], and then the effective Lagrangian
 contains a meson field as a bilocal field. At first, we take the
 light-cone gauge $ A^+=0$.  The remaining gauge field
 $A^-$ induces non-local
 four-fermi interaction with light-cone distance $r^-$
 at equal light-cone time $x^+$. In this interaction term, we
 introduce the bilocal auxiliary fields using the path-integral,
 and this field corresponds to the meson. The Fourier
 transformation of the bilocal auxiliary fields determines the parton
 momentum distribution function. From the effective Lagrangian, we
 obtain the 't Hooft equation for the bilocal auxiliary fields.

 The paper is organized as follows. In section 2, we briefly review
 QCD$_2$ with the light-cone gauge. In section 3, we derive the
 large-$N_c$ effective Lagrangian, and we decompose the auxiliary fields
 into the vacuum expectation value and its fluctuation field. This
 vacuum expectation value of the auxiliary fields is obtained by the
 effective potential. In section 4, we discuss the effective potential
 and the vacuum expectation value, which is
 obtained from the gap equation and determines the dynamical quark
 mass. In section 5, we derive the 't Hooft equation from our effective
 Lagrangian. We conclude the paper in section 6, and we summarize our
 discussion.

\section{$\mbox{SU}(N_c)$ QCD}

 An $\mbox{SU}(N_c)$ gauge theory is defined by the Lagrangian  
\begin{equation}
  {\cal L}=\overline{q}_i ( i\gamma ^{\mu} D_{\mu}-m_i )\ q_i 
          -\frac{1}{4}\ F_{a\mu\nu}F_{a}^{\mu\nu},
\end{equation}
 where $i$ denotes the flavour ($i=1,\cdots ,N_f$) and $a$ denotes the
 adjoint representation of colour ($a=1,\cdots ,N_c^2-1$). 
 $F_{a\mu\nu}$ is the field-strength tensor
 $F_{a\mu\nu}=\partial_\mu A_{a\nu}-\partial_\nu
 A_{a\mu}-gf_{abc}A_{b\mu}A_{c\nu}$, and the covariant derivative is
 defined as $D_{\mu}=\partial_\mu +igt^a A_{a\mu}$. The fermion field
 $q_i$ is the fundamental representation of SU$(N_c)$, and in two
 dimensions this field is a two-component spinor. The light-cone coordinate
 is defined by $x^+=x^0+x^3,\ x^-=x^0-x^3$.\ We employ a specific
 representation $\gamma^{\pm}=\sigma_1\pm i\sigma_2$. In this
 representation, the spinor is given by $q^T=(q_-,\ q_+)$, where the
 lower component~(upper component) $q_+~(q_-)$ is a projection of
 $\Lambda^{\pm}=\frac{1}{4}\gamma^{\mp}\gamma^{\pm}$. The light-cone gauge
 is given by $A^+=0$. In this gauge there are no ghosts, and the remaining
 gauge field is only $A^-$. Then the Lagrangian is given by
\begin{eqnarray}
      {\cal L}\!&=&\!
q_+^{i\dag}~i\partial^-~ q^i_+ 
+ q_-^{i\dag}~ i\partial^+~  q^i_- 
    -m_i \left( q_+^{i\dag}~  q^i_- + q_-^{i\dag}
          ~ q^i_+\right) \nonumber \\ 
       && \hspace*{39mm}-~ g q_+^{i\dag}~  t^a A_a^- ~ q^i_+ 
           +\frac{1}{8}\left( \partial^+ A_a ^-\right)^2.
\end{eqnarray} 
Integrating over $A^-$, this gauge field induces a non-local four-fermi
interaction:
\begin{eqnarray}
      {\cal L}\!&=&\!
  q_+^{i\dag}~ i\partial^-~  
              q^i_+ + q_-^{i\dag}~  i\partial^+~ q^i_-
                       -m_i \left( q_+^{i\dag}~ q^i_- 
          + q_-^{i\dag}~ q^i_+\right)\nonumber \\
       &&  \hspace*{39mm}  +~ 2 g^2~\left(q_+^{i\dag}~ t^a~ q^i_+ \right)
\frac{1}{\partial^{+2}} \left( q_+^{j\dag}~  t^a~ q^j_+\right),
\end{eqnarray}
where the inverse operator $\frac{1}{\partial^{+2}}$ is defined by
~\cite{coleman2,einhorn}
\begin{equation}
    \frac{1}{\partial^{+2}}q_+(x^-)
    =\frac{1}{8}\int^\infty _{-\infty} dy^-
                    |x^--y^-|q_+(y^-).
\end{equation}
The equation of motion for $q^{i\dag }_+$ is given as $q^i_- =\frac{m_i
}{i\partial ^+}q^i_+$, and then the field $q^i_-$ is the constrained
field because at any $x^+$ it is determined by $ q^i_+$. The action is
given in terms of the one-component spinor $ q^i_+$ as: 
\begin{eqnarray}
      \lefteqn{S= 
                \frac{1}{2}\int\! dx^- dx^+ ~ 
                          q_+^{i\dag} 
                        \left(  i\partial^- 
               - ~ \frac{m_i^2 }{i\partial ^+}\right) 
                                         q^i_+}\hspace{10mm}\nonumber \\
       &&\hspace*{5mm}+\frac{g^2}{8}\int dx^-dx^+ dy^-
                        \left(q_+^{i\dag}(x^-,x^+)~ 
                    t^a  ~ q^i_+ (x^-,x^+) \right)\nonumber \\
       && \hspace*{30mm}\times|x^--y^-| 
                         \left( q_+^{j\dag} 
                 (y^-,x^+)~ t^a  ~ q^j_+(y^-,x^+)\right).   
\end{eqnarray}
The non-local four-fermi interaction with linear potential induces the
bound state of quark and antiquark.

\section{The large-$N_c$ effective action with the auxiliary fields}

We expand the Lagrangian by $1/N_c$, with $g^2N_c$ fixed.
We discuss the theory in the leading order of $1/N_c$ expansion. 
At first we rearrange the non-local four-fermi interaction 
by using the identity
$(t_a)_{\alpha\beta}(t_b)_{\gamma\delta}=
-\frac{1}{2N_c}\delta_{\alpha\beta}\delta_{\gamma\delta}
+\frac{1}{2}\delta_{\alpha\delta}\delta_{\gamma\beta}$.
In the leading order only the colour singlet part of the quark pair
contributes. Then the effective action in the large-$N_c$ limit
is given by  
\begin{eqnarray}
      S^{{\rm Large}N_c}
         \!&=&\!\frac{1}{2}\int dx^-dx^+ 
          q^{\dag i}_+\left(i\partial ^-
          -\frac{m^2}{i\partial ^+}\right)q^i_+
                                    \nonumber \\
     &&\hspace*{-5mm}-\frac{g^2}{16}\int dx^-dx^+dy^-
         \left(q^{\dag i}_+ (x)q^j_+(y)\right) |x^--y^-|
          \left(q^{\dag j}_+ (y)q^i_+(x)\right),        
\end{eqnarray}
where the quark pair in the interaction term is a colour singlet,
and this quark pair is non-local in the light-cone space, 
that is  $q^{\dag i}_+(x)q^j_+(y)\equiv
q^{\dag i}_{+\alpha}(x^-,x^+)~q^j_{+\alpha}(y^-,x^+)$. 
Here we ignore next-to-leading terms. 
The partition function of this theory is written as
\begin{equation}
     Z^{{\rm Large}N_c}
      =\int \prod_{i}[dq^{\dag i}_+][dq^i_+]
        \exp\left\{iS^{{\rm Large} N_c}\right\}.
\end{equation}
In order to treat the singlet of quark pair as a dynamical variable, we
introduce bilocal auxiliary fields. After integrating over quark fields,
the bilocal auxiliary field corresponds to the singlet of the quark pair.
At first we multiply the partition function by a constant factor of
$\delta$-functions:
\begin{equation}
     Z^{{\rm Large}N_c}
    =\int \prod_{i,j}[dq^{\dag i}_+][dq^i_+][ds^{ij}]
      \exp\left\{iS^{{\rm Large} N_c}\right\}
       \prod\limits_{i,j} 
        \delta\left( s^{ij}-q^{\dag j}_+ (x)q^i_+(y)\right),
\end{equation} 
where we introduce the auxiliary fields $s^{ij}$, which correspond to
$q^{\dag j}_+ (x)q^i_+(y)$. These $\delta$-functions are defined
with the additional bilocal fields $\sigma$: 
\begin{eqnarray}  
     \lefteqn{ \prod\limits_{i,j} 
      \delta \left( s^{ij}(y,x)-q^{\dag j}_+ (x)q^i_+(y)
                                   \right)}\hspace{15mm}\nonumber \\
              &\propto& \int\! \prod\limits_{i,j}[d\sigma _{ij}]~
         \exp \left\{ \frac{i}{2}\int\! dx^- dx^+ dy^- 
         ~\sigma _{ji}(x^-,y^-;x^+)\right.\nonumber \\
    &&\hspace*{40mm}  
       \left.\times (s^{ij}(y,x)-q^{\dag i}_+ (x)q^j_+(y))\right\}.
\end{eqnarray}
The partition function with the bilocal auxiliary fields is given by
\begin{equation}
   Z^{{\rm Large}N_c}
       =\int \prod_{i,j}[dq^{\dag i}_+][dq^i_+][ds^{ij}][d\sigma_{ij}]
         \exp\left\{iS^{{\rm Large} N_c}+iS^{\rm Aux}\right\},
\end{equation}
where the action $S^{\rm Aux}$ is written as 
\begin{equation}
        S^{\rm Aux}
            =\frac{1}{2}\int\! dx^- dx^+ dy^- 
             ~\sigma _{ji}(x^-,y^-;x^+)
               \left(s^{ij}(y,x)-q^{\dag j}_+ (x)q^i_+ (y)\right).         
\end{equation}
In order to consider the correspondence between the bilocal fields
$\sigma_{ij}$ and quarks $q^i_+$, varying $Z^{{\rm Large}N_c}$ by $
s^{ij}(x,y)$, we get the relation  
\begin{equation}
       \langle \sigma _{ij}(x^-,y^-;x^+) \rangle
       =\langle \frac{g^2}{8}|x^--y^-|q^{\dag j}_+(y)q^i_+(x) \rangle.
\end{equation}
This equation tells us that the auxiliary fields are of order $g^2$ and
that these fields correspond to quark pairs with gluons; they would then
be identified as mesonic dynamical variables. This will be mentioned later. 
Integrating over auxiliary fields $s^{ij}$ and quarks $q^i_+$, the 
Lagrangian is given by  
\begin{eqnarray}
    \hspace*{-12mm}\lefteqn{ S^{{\rm Large}N_c}
       =-iN_c\ln{\rm Det}~M(x,y)}\hspace{39mm} \nonumber \\
              &&\hspace*{-28mm} +\int dx^- dx^+ dy^- 
                 \sigma _{ij}(y^-,x^-;x^+)
                  \frac{1}{g^2 |x^--y^-|}\sigma_{ji}(x^-,y^-;x^+),
\end{eqnarray}
where the matrix $M$ is defined by
\begin{equation}
      M(x,y)\equiv \left[\ \delta_{ji}
           \left(i\partial ^- 
             -\frac{m^2}{i\partial ^+} \right)\ \delta^2(x-y)
              -\sigma_{ji}(x,y)\ \right],
\end{equation}
with
\begin{equation}
        \sigma_{ij}(x,y)
         \equiv 
          \sigma_{ij}(x^-,y^-;x^+)\delta(x^+-y^+).
\end{equation}
The vacuum expectation value
of these bilocal auxiliary fields $\sigma_{ij}$ is obtained by the
effective potential. From the charge conservation, the diagonal part of
$\sigma_{ij}$ takes non-zero vacuum expectation value, and the other is zero:
\begin{eqnarray}
      \langle \sigma_{ij}(x^-,y^-;x^+) \rangle
     \!&=&\!\left\{\begin{array}{lll}
           v_i\ (x^--y^-)&\ {\rm if}&  i=j,\\
                             0&\ {\rm if}& i \neq j.
                           \end{array}\right.
\end{eqnarray}
The bilocal auxiliary fields are expanded into the vacuum expectation
values and the fluctuation, 
\begin{equation}    
     \sigma_{ij}(x^-,y^-;x^+)
     \longrightarrow 
    v_i\ (x^--y^-)\ \delta_{ij}
                  +\sigma_{ij}(x^-,y^-;x^+).
\end{equation}
Finally we arrive at the large-$N_c$ effective action: 
\begin{equation}
     S^{{\rm Large}N_c}=S^{(0)}+S^{(2)},\label{ac}
\end{equation}
where
\begin{eqnarray}
     S^{(0)}
     \!&=&\!-iN_c{\rm TrLn}~S^{-1}_F (y-z)\nonumber\\
            &&\hspace*{15mm} +\int\! dx^- dx^+ dy^- 
               \frac{v_i(y^--x^-)v_i(x^--y^-) }{g^2 |x^--y^-|},\label{ac0}\\
 S^{(2)}\!&=&\!\frac{iN_c}{2} {\rm Tr}[S_F(z-u)\Sigma(u,x)]^2 \nonumber\\
               &&\hspace*{15mm}+\int\! dx^- dx^+ dy^- 
            \frac{\sigma _{ij}(y^-,x^-;x^+) 
                   \sigma_{ji}(x^-,y^-;x^+)}{g^2 |x^--y^-|}.\label{ac2}
\end{eqnarray}	
Here the subscript of $S$ denotes the degree of $\sigma$.
The first-degree action $S^{(1)}$ vanishes if we use the gap equation of
$v^\dag _i $, which will be discussed in the next section. In this
Lagrangian, the matrices $S^{-1}_F$ and $\Sigma$ are defined by 
\begin{equation}
        S^{-1}_F(x-y)
            ={\rm diag}\left( \left(i\partial ^-
             -\frac{m^2_i}{i\partial ^+} \right)\ \delta^2 (x-y)
           -v_i\ (x^--y^-)\delta(x^+-y^+)\right),\label{ipro}
\end{equation}  
and
\begin{equation}
       \Sigma(x,y)
       =\Bigl(\ \sigma_{ij}(x,y)\ \Bigl).
\end{equation} 
Here we ignore the higher-order terms of the $1/N_c$ 
expansion in Eq. (\ref{ac}).

\section{The effective potential and the gap equation}

 In this section, we discuss the behaviour of the vacuum expectation value
 of the bilocal auxiliary field by considering the leading order
 $S^{(0)}$ solely. We give the momentum space description of
 $S^{(0)}$. Then we can easily extract the space-time volume from the
 effective action. The first term in Eq. (\ref{ac0}) is  
\begin{eqnarray}
        \lefteqn{{\rm TrLn}~S^{-1}_F(y-x)} \nonumber\\                         &=& \int dx^+ dx^- dy^+ dy^- \ \delta^2(y-x)
          \int\! \frac{d^2p}{(4\pi)^2}~ e^{ip\cdot (y-x)} 
           {\rm ln}\det S^{-1}_F(p) \nonumber\\
         &=& \delta^2(0)\int\! dp^+dp^-
             ~ \sum _i {\rm ln}
                \left(p^--\frac{m^2_i}{p^+}-v_i(p^+)\right),
\end{eqnarray}
where $S^{-1}_F(p)$ and $v(p^+)$ are the Fourier transformation of
Eq. (\ref{ipro}) and $v_i(x^-)$:
\begin{eqnarray}
       S^{-1}_F(p) 
              \!&=&\!{\rm diag}\left(p^-
               -\frac{m^2_i}{p^+}-v_i(p^+)\right),\\
    v_i(x^-)
           \!&=&\!\int\! \frac{dp^+}{4\pi}
          e^{-\frac{i}{2}p^+x^-}v_{i}(p^+). 
\end{eqnarray} 
The second term in Eq. (\ref{ac0}) is
\begin{eqnarray}
        \lefteqn{\int dx^+ dx^- dy^- 
         \frac{v_i (y^--x^-)v_i(x^--y^-)}{g^2 |x^--y^-|}}
          \nonumber\\
    &&\hspace{5mm}=~4\pi\delta(0)\int dx^- dy^- 
         \frac{v_i(y^--x^-)v_i(x^--y^-)}{g^2
                  |x^--y^-|}\nonumber\\
    &&\hspace{5mm}=~\delta^2(0)\int dp^+ dk^+ v_i
                  (p^+)G(p^+-k^+)v_i(k^+),
\end{eqnarray}
where we define the kernel $G$: 
\begin{equation}
          \frac{1}{g^2|x^-|}
          =\int \frac{dk^+}{4\pi}
        e^{-\frac{i}{2} k^+x^-}G(k^+).
\end{equation}
The effective potential is defined by $S^{(0)}=-\delta^{2}(0)V_{\rm eff}$, then
\begin{eqnarray}
           V_{\rm eff}
           &=&iN_c\sum_i \int dp^+dp^-
                  {\rm ln}\left(p^--\frac{m^2_i}{p^+}
                   -v_i(p^+)\right)\nonumber\\
           &&\hspace*{30mm}-\int dp^+ dk^+ v_i (p^+)G(p^+-k^+)v_i(k^+).
\end{eqnarray}
The behaviour of the vacuum expectation value $v_i (p^+)$ is determined
by the gap equation, which is given by varying the effective potential
with respect to $v_i (p^+)$:
\begin{eqnarray}
     \frac{\delta V_{\rm eff}}{\delta v_{i} (p^+)}
     &=&-iN_c \int dp^- \frac{1}{p^--\frac{m^2_{i}}{p^+}
        -v_{i}(p^+)+i\epsilon~{\rm sgn}(p^+)}\nonumber\\
     &&\hspace*{43mm} - 2 \int dk^+ G(p^+-k^+)v_{i}(k^+)\nonumber\\
     &=&0.
\end{eqnarray}
Multiplying these equations by the inverse of $G(p^+)$, the gap equation
is written as
\begin{equation}
     v_i(p^+)=\frac{i N_c g^2}{4 \pi ^2}
               \int dk^+dk^- 
                \frac{\cal P}{(p^+-k^+)^2}
                 \frac{1}{k^--\frac{m^2_i}{k^+}
       -v_i(k^+)+i\epsilon~{\rm sgn}(k^+)}.\label{gap} 
\end{equation}
The symbol ${\cal P}$ means the principal value of
$(p^+-k^+)^{-2}$. This equation is equivalent to the Schwinger--Dyson
equation. The vacuum expectation value gives the self-energy of the quark
and the dynamical quark mass. Performing the integration in
Eq. (\ref{gap}), we arrive at $v_i (p^+)=-\frac{N_c
g^2}{2\pi}~\frac{1}{p^+}$. However we will not use this explicit
expression in the derivation of the 't Hooft equation discussed in the
next section.    
\begin{figure}[t]
\begin{center}
\leavevmode
\psfig{file=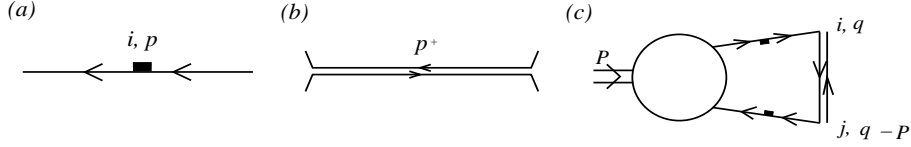,width=12cm}
\caption{Feynman rule:~(a) quark propagator \small{$S_F^i(p)=\left(p^--\frac{m^2_i}{p^+}-v_i(p^+)+i\epsilon~{\rm sgn}(p^+)\right)^{-1}$}, ~which is the diagonal part of $S_F(p)$, (b) gluon propagator, with vertices $G^{-1}(p^+)=-{\mbox {\scriptsize $8g^2$}}\frac{{\cal P}}{p^{+2}}$ ; and (c) bilocal field $\sigma_{ij}(q^+,q^+-P^+:P^-)$.}
\label{feyn} 
\end{center}
\end{figure} 

\section{ The 't Hooft equation}

\begin{figure}[t]
\begin{center}
\leavevmode
\psfig{file=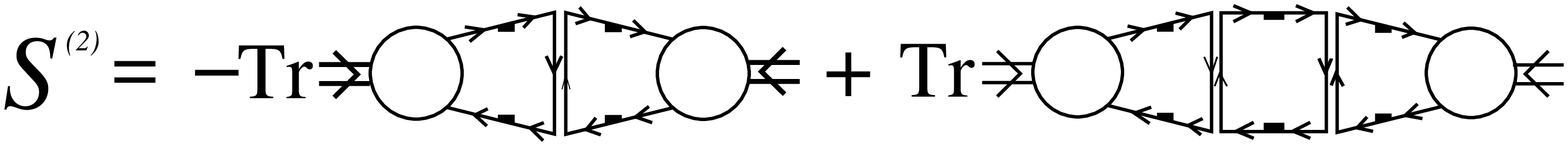,width=12cm} 
\caption{ Diagrammatic representation of the effective action (\ref{effac}).}
\label{action}
\end{center}
\end{figure}

\begin{wrapfigure}{l}{6.6cm}
\leavevmode
\psfig{file=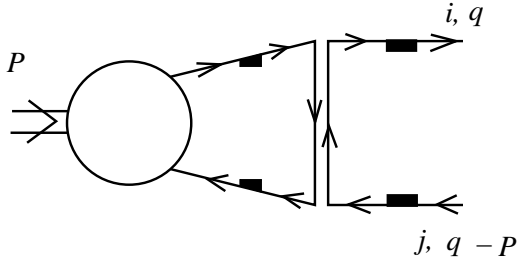,width=6.8cm}
\caption{Diagrammatic representation of $\psi_{ij}(q^+, q^+-P^+;P^-)$.}
\label{psi}
\end{wrapfigure}

Our bilocal auxiliary field is supposed to be the mesonic field in the
large-$N_c$ theory. In two dimensional large-$N_c$ QCD, there is a
well-known bound-state equation called the 't Hooft equation. If
our bilocal auxiliary field is a bound state, it should satisfy this 
equation. This equation was first derived in \cite{thooft}, and 
the mesonic field was introduced as a solution of the
ladder Bethe--Salpeter equation. As we will show, our effective theory
contains the bilocal field as a meson, which satisfies the
't Hooft equation. In this section we derive the 't Hooft equation from
our effective theory. At first we give the definition of 
the Fourier transformation, which, for 
$\sigma_{ij}(x^-,y^-;x^+)$, is defined by
\begin{equation}
     \sigma_{ij}(x^-,y^-;x^+)
   =\int\!\frac{dP^-}{4\pi} \frac{dp^+}{4\pi}
       \frac{dk^+}{4\pi}
                 e^{-\frac{i}{2}P^- x^+}
                 e^{-\frac{i}{2}(p^+x^--k^+y^-)}
                 ~\sigma_{ij}(p^+,k^+;P^-), 
\end{equation} 
or in the variables $p^+\rightarrow q^+$ and $k^+\rightarrow q^+-P^+$,
\begin{eqnarray}
     \lefteqn{\sigma_{ij}(x^-,y^-;x^+)
       =\int\! \frac{dP^-}{4\pi} \frac{dq^+}{4\pi}
           \frac{dP^+}{4\pi}
        e^{-\frac{i}{2}P^- x^+}
        e^{-\frac{i}{2}(q^+(x^--y^-)+P^+y^-)}}\hspace{25mm}\nonumber\\
        && \hspace*{55mm}\times\sigma_{ij}(q^+,q^+-P^+;P^-),
\end{eqnarray}
where this $P^+$ is the total momentum and $q^+$ is the momentum
carried by the quark (see Fig. \ref{feyn}). 
The quark propagator is given by
\begin{eqnarray}
    S_F (p)&=&\int\! dx^+ dx^- e^{ip\cdot x}S_F(x)\nonumber \\
            &=&{\rm diag}\left(
                \frac{1}{p^--\frac{m^2_i}{p^+}
                -v_i (p^+)+i\epsilon~{\rm sgn}(p^+)}
                \right).
\end{eqnarray}
The Feynman rule is shown in Fig. \ref{feyn}.
The momentum-space description of the action (\ref{ac2}) is given by
\begin{eqnarray}
     S^{(2)} 
       \!&=&\!\frac{iN_c}{2}{\rm tr} \int\! 
           \frac{dP^- dP^+ dq^- dq^+}{(4\pi)^4}
             ~S_F (q^+,q^-)
           ~\Sigma(q^+,q^+-P^+;P^-) \nonumber\\ 
       &&\hspace*{35mm} 
          \times S_F(q^+-P^+,q^--P^-) 
          ~\Sigma(q^+-P^+,q^+;-P^-)
            \nonumber\\ 
       && +{\rm tr}\int\! \frac{dP^-dP^+ dq'^+ dq^+}{(4\pi)^4}
           ~\sigma(q'^+,q'^+-P^+;P^-)\nonumber\\ 
       &&\hspace*{35mm}\ \times G(q'^+-q^+)~\sigma(q^+-P^+,q^+;-P^-).
                     \label{effac}
\end{eqnarray} 
This action is depicted in Fig. \ref{action}. 
The Bethe--Salpeter equation is derived by varying the $S^{(2)}$ with
respect to the bilocal field $\sigma_{ji}(q^+-P^+,q^+;-P^-) $ 
\begin{eqnarray}
         \lefteqn{(4\pi)^4
          \frac{\delta S^{(2)}}
                {\delta \sigma_{ji}(q^+-P^+,q^+;-P^-)}}\hspace{10mm}\nonumber\\
       &=&iN_c\int dq^- S^i_F(q^+,q^-)\sigma_{ij}(q^+,q^+-P^+;P^-)
                 S^j_F(q^+-P^+,q^--P^-)\nonumber\\
       &&+2\int dq'^+ G(q^+-q'^+)                         
            \sigma_{ij}(q'^+,q'^+-P^+;P^-)\nonumber\\
       &=&0 .\label{bseq}
\end{eqnarray}
This equation is depicted in Fig. \ref{bs}.
We multiply the Bethe--Salpeter equation by the kernel $G^{-1}$, which
is written as a simple form:
\begin{eqnarray}
         \lefteqn{\sigma_{ij}(q^+,q^+-P^+;P^-)}\hspace{8mm}\nonumber\\ 
       &=&\frac{iN_c g^2}{4\pi^2} 
           \int dk^- dk^+ \frac{{\cal P}}{(q^+-k^+)^2}\nonumber\\
    &&\hspace*{12mm}\times\frac{1}{k^--\frac{m_i^2}{k^+}-v_i(k^+) +i\epsilon~
       {\rm sgn}(k^+)}
        \sigma_{ij}(k^+,k^+-P^+;P^-)\nonumber\\
    &&\hspace*{12mm}\times\frac{1}{k^--P^-
         -\frac{m_j^2}{k^+-P^+}-v_j(k^+-P^+)
         +i\epsilon~{\rm sgn}(k^+-P^+)}.\label{bseq2}
\end{eqnarray}
In Eq. (\ref{bseq2}) the variable $k^-$ is only included in the quark
propagators, and we can easily perform the integration over $k^-$. If the
condition ${\rm sgn}(k^+)=-{\rm sgn}(k^+-P^+)$ is satisfied, the
integration over $k^-$ gives the finite contribution.
If $P^+$ is positive\footnote{In spite of the sign of $P^+$, we arrive at
the same expression Eq.(\ref{bseq3}). In the light-cone quantization,
$P^+$ is positive, because of the dispersion relation and the
positive-definite light-cone energy.}, the integration region of $k^+$ is
restricted as follows:
\begin{eqnarray}
           0\ <\ k^+\ <\ P^+.
\end{eqnarray}
Then the Bethe--Salpeter equation is written as      
\begin{eqnarray}
       \lefteqn{\sigma_{ij}(q^+,q^+-P^+;P^-)}&&\hspace{10mm}\nonumber\\ 
      \hspace*{-30mm} \!&=&\! -\frac{N_c g^2}{2\pi}\!
          \int_0^{P^+}\!dk^+
           \frac{{\cal P}}{(q^+-k^+)^2}
           \frac{\sigma_{ij}(k^+,k^+-P^+;P^-)}
       {P^--(\frac{m_i^2}{k^+}+\frac{m_j^2}{P^+-k^+})
                -v_i(k^+)+v_j(k^+-P^+)}.\label{bseq3}
\end{eqnarray}  
We define the field $\psi_{ij}$ by the field $\sigma_{ij}$ and two
quark propagators:
\begin{figure}[t]
\begin{center}
\leavevmode
\psfig{file=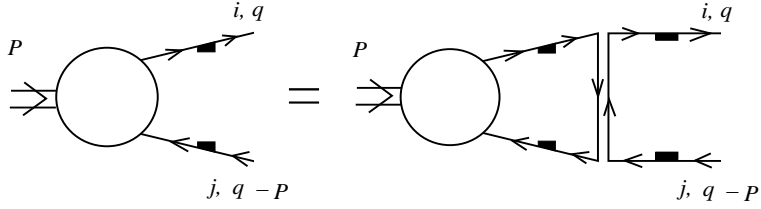,width=10cm}
\caption{Diagrammatic representation of Eq. (\ref{bseq}).}
\label{bs}
\end{center}
\end{figure} 
\begin{equation}
     \psi_{ij}(q^+,q^+-P^+;P^-)\equiv 
        \frac{\sigma_{ij}(q^+,q^+-P^+;P^-)}
   {P^--(\frac{m_i^2}{q^+}+\frac{m_j^2}
       {P^+-q^+})-v_i(q^+)+v_j(q^+-P^+)}.
    \label{psieq}
\end{equation}
This field $\psi_{ij}$ is depicted in Fig. \ref{psi}.
 Using the field $\psi_{ij}$, the Bethe--Salpeter equation is rewritten as 
\begin{eqnarray}
          \lefteqn{\left[\ P^-- \left(\frac{m_i^2}{q^+}
                   +\frac{m_j^2}{P^+-q^+}\right)
              -v_i(q^+)+v_j(q^+-P^+)\right]
             \psi_{ij}(q^+,q^+-P^+;P^-)}
              \hspace{30mm}\nonumber\\
          \hspace*{-27mm} &=&-\frac{N_c g^2}{2\pi}{\cal P} 
                \int_0^{P^+}\! dk^+ 
                ~\frac{\psi_{ij}(k^+,k^+-P^+;P^-)}{(q^+-k^+)^2}.
\end{eqnarray}
The bound-state mass $M$ and the momentum fraction variable $x$ are
defined as
\begin{eqnarray}
            M^2\equiv P^+P^-,\ x\equiv \frac{q^+}{P^+}.
\end{eqnarray}
The field $\psi_{ij}$ is rewritten in terms of the momentum fraction $x$,
\begin{eqnarray}
           \psi_{ij}(xP^+)=\psi_{ij}(q^+,q^+-P^+;P^-).
\end{eqnarray}
\\
Using these variables, the Bethe--Salpeter equation is written as
\vspace{3mm}
\begin{eqnarray}
        \lefteqn{ \left[\ M^2-\left(\frac{m_i^2}{x}+\frac{m_j^2}{1-x}
                \right)
               -P^+\ v_i(q^+)+P^+\ v_j(q^+-P^+)\right]
                    \psi_{ij}(xP^+) }
                     \hspace{50mm}\nonumber\\
                  &=&-\frac{N_c g^2}{2\pi} {\cal P}
                       \int_0^{1} dy 
                        \frac{\psi_{ij}(yP^+)}{(x-y)^2}.
\end{eqnarray}
The difference between the two vacuum expectation values is easily
calculated by using the gap equation Eq. (\ref{gap}),
\begin{eqnarray}
      v_i(q^+)-v_j(q^+-P^+)=\frac{N_c g^2}{2\pi} 
          \int_0^{P^+} dk^+\frac{{\cal P}}{(q^+-k^+)^2}.
\end{eqnarray}
At last, the 't Hooft equation is derived:
\begin{equation}
         \left[\ M^2-\left(\frac{m_i^2}{x}
         +\frac{m_j^2}{1-x}\right)\ \right]
           \psi_{ij}(xP^+)
          =-\frac{N_c g^2}{2\pi} 
             {\cal P}\int_0^{1} dy 
              \frac{\psi_{ij}(yP^+)
             -{\psi_{ij}(xP^+)}}{(x-y)^2}.
\end{equation}
In our large-$N_c$ effective theory, this equation is derived using a
bilocal auxiliary field. This shows that the auxiliary field is a
collective field, corresponding to the bound state.

\section{Summary}

 We construct the large-$N_c$ QCD$_2$ effective theory of the bilocal 
 auxiliary fields. In this theory we introduce the bilocal auxiliary 
 field as a colour singlet, which is bilocal in the light-cone space
 and local in the light-cone time. This non-locality means the
 distance between the constituent quarks. From the consideration of the
 't Hooft equation, we identify the bilocal field as the parton momentum
 distribution function, or the meson field. Original derivation of the 't
 Hooft equation was based on the ladder Bethe--Salpeter equation, and
 the meson field is introduced as the solution of this equation. 
 From our point of view, there is a meson theory that is reduced from 
 large-$N_c$ QCD$_2$, and from this theory we derive the 't Hooft equation. 
 In our approach the vacuum expectation value of the bilocal field gives 
 the dynamical quark mass, and then it corresponds to the self-energy of 
 the quark. We may suppose the result of the gap equation contradicts the
 triviality of the light-cone vacuum. In this respect, our auxiliary 
 field is bilocal and indeed has a vacuum expectation value. 
 The result implies that the light-cone vacuum is not trivial. 
\vspace{7mm}\\
{\Large \bf Acknowledgements}\vspace{4mm}\\
We like to thank S. Fukae for collaboration at the early stage. We also
like to thank V. Gusynin, K. Harada, M. Harada, M. Hirata and Y. Yamamoto
for discussions. T. M. likes to thank the PQCD working group members for
their encouragement. The work of T. M. is supported by Grand-in Aid for
Special Project Research (Physics of CP violation) under the 
Grant NO. 12014211 and by Fellowship for
Japanese scholar and researcher in abroad from the Ministry of
Education, Science, and Culture of Japan.


\end{document}